\documentclass[a4paper,fleqn]{cas-sc}

\usepackage[authoryear,longnamesfirst]{natbib}

\usepackage{xcolor}
\usepackage[normalem]{ulem}

\begin{document}

\let\WriteBookmarks\relax
\def\floatpagepagefraction{1}
\def\textpagefraction{.001}

\shorttitle{First-order phase transitions in the three-dimensional Blume-Capel ferromagnet}

\shortauthors{D. Mataragkas et al.}
 
\title {First-order phase transitions in the three-dimensional Blume-Capel ferromagnet}                      
\author[1]{D. Mataragkas} [orcid=0009-0008-9058-8923]
\credit{Software, Methodology, Writing--Original draft preparation}
\affiliation[1]{organization={School of Mathematics, Statistics and Actuarial Science, University of Essex},
city={Colchester},
postcode={CO4 3SQ}, 
country={United Kingdom}}

\author[1]{A. Vasilopoulos} [orcid=0000-0002-9415-8970]
\credit{Software, Methodology, Writing--Original draft preparation}

\author[2]{Z.D. Vatansever} [orcid=0000-0002-7726-5168]
\credit{Methodology, Writing--Editing}
\affiliation[2]{organization={Department of Physics, Dokuz Eyl\"{u}l
University},
city={Izmir},
postcode={TR-35160}, 
country={Turkey}}

\author[3]{Y. Kim} [orcid=0009-0003-9347-2081]
\credit{Validation, Methodology}

\author[3]{D.-H. Kim} [orcid=0000-0003-3896-2136]
\credit{Conceptualization, Validation, Methodology, Writing–Editing}
\affiliation[3]{organization={Department of Physics and Photon Science, Gwangju Institute of Science and Technology},
city={Gwangju},
postcode={61005}, 
country={Republic of Korea}}
\cormark[1]
\ead{dongheekim@gist.ac.kr}
\cortext[cor1]{Corresponding author}

\author[1]{N.G. Fytas} [orcid=0000-0002-9428-1709]
\credit{Conceptualization, Validation, Methodology, Writing–Editing}
\cormark[2]
\ead{nikolaos.fytas@essex.ac.uk}
\cortext[cor1]{Corresponding author}

\begin{abstract}
We investigate first-order phase transitions in the three-dimensional Blume-Capel ferromagnet on the simple cubic lattice by combining multicanonical simulations with two-parameter Wang-Landau sampling. The generalized-ensemble approach enables a detailed characterization of the coexistence region over a broad range of temperatures and crystal-field couplings, extending from the vicinity of the tricritical point deep into the first-order regime. We analyze the finite-size scaling behavior of thermodynamic observables, with particular emphasis on the energy probability density function, free-energy barriers, and interfacial properties. The evolution of the double-peaked energy distributions reveals a gradual crossover from weak to strong first-order behavior as the temperature is lowered. From the scaling of the free-energy barrier we determine the interface tension along the coexistence line and characterize its approach to the tricritical region through its vanishing behavior. In parallel, a field-mixing analysis based on the joint density of states obtained from two-parameter Wang-Landau simulations is employed to locate first-order transition points and probe tricritical behavior. While this approach has been highly successful in two-dimensional realizations of the Blume-Capel model, we find that in three dimensions its practical implementation becomes increasingly sensitive in the vicinity of the tricritical region, where the shallow structure of the relevant scaling variable distribution limits the ability to resolve coexistence conditions for the system sizes currently accessible. These results delineate the range of applicability of the method in three dimensions and provide a consistent picture of the first-order regime of the model. 
\end{abstract}

\begin{keywords}
Blume-Capel model \sep first-order transitions \sep tricriticality  \sep Monte Carlo simulations \sep finite-size scaling
\end{keywords}

\maketitle

\section{Introduction}
\label{sec:Intro}

First-order phase transitions constitute a fundamental class of collective phenomena in statistical physics, characterized by phase coexistence, latent heat, and the emergence of free-energy barriers separating competing thermodynamic states~\cite{binder87}. Despite their apparent simplicity, first-order phase transitions exhibit a rich phenomenology, including metastability and phase coexistence, together with characteristic finite-size effects such as double-peaked energy distributions and nontrivial scaling behavior~\cite{binder84}. Understanding how these features evolve across different regions of a phase diagram remains an active area of research, particularly in systems where first-order and continuous transitions are connected through a tricritical point.

Among the lattice models displaying such behavior, the Blume-Capel model~\cite{blume,capel} occupies a prominent position as one of the simplest and most extensively studied realizations of tricriticality. The model provides a paradigmatic framework for investigating the interplay between continuous and discontinuous phase transitions and has served for several decades as a benchmark for analytical and numerical studies of critical phenomena~\cite{lawrie}. Beyond its theoretical importance, the model is relevant to a variety of physical systems, including multicomponent fluids, ternary alloys, and $^{3}$He--$^{4}$He mixtures~\cite{lawrie}. More recently, variants of the model have been employed in the study of ferrimagnetic materials~\cite{selke-10}, wetting and interfacial adsorption phenomena~\cite{fytas13}, as well as in investigations of the scaling properties of partition-function zeros~\cite{leila} and energy-distribution zeros~\cite{macedo24}.

In the absence of an external magnetic field, the model is described by the Hamiltonian
\begin{equation}\label{eq:Hamiltonian}
\mathcal{H}
=-J\sum_{\langle xy\rangle}\sigma_{x}\sigma_{y}+\Delta\sum_{x}\sigma_{x}^{2}
=E_{J}+\Delta E_{\Delta},
\end{equation}
where the spin variables $\sigma_x$ assume the values $-1$, $0$, and $+1$, the summation $\langle xy\rangle$ runs over all nearest-neighbor pairs, and $J>0$ denotes the ferromagnetic exchange interaction, which is set to be unity throughout this paper. The crystal-field coupling $\Delta$ controls the concentration of vacancies corresponding to the state $\sigma_x=0$. The decomposition of the Hamiltonian into the bond-energy contribution $E_J$ and the crystal-field contribution $E_\Delta$ will prove particularly useful in the generalized-ensemble simulations employed in this work. In the limiting case $\Delta\rightarrow -\infty$, vacancies are completely suppressed and the model reduces to the conventional spin-$1/2$ Ising ferromagnet.

The phase diagram of the Blume-Capel model in the $(\Delta,T)$ plane exhibits a rich structure. The phase boundary separates a ferromagnetic phase, characterized by long-range ordering of the $\pm 1$ spin states, from a paramagnetic phase. Depending on the values of temperature and crystal field, the latter may correspond either to a fully disordered spin configuration or to a dilute gas of $\pm 1$ spins embedded in a vacancy-dominated background. For weak crystal fields and sufficiently high temperatures, the transition between the two phases is continuous and belongs to the Ising universality class. Increasing the crystal field and lowering the temperature eventually leads to a discontinuous transition accompanied by phase coexistence and latent heat~\cite{blume,capel}. The continuous and first-order segments of the phase boundary meet at a tricritical point $(\Delta_{\rm t},T_{\rm t})$, rendering the Blume-Capel model a prototypical system for the study of tricritical behavior~\cite{lawrie,deserno97,mataragkas25a}. Throughout the present work, we adopt the tricritical-point estimate of Deserno~\cite{deserno97} as a reference. We note, however, that a slightly different estimate, $\Delta_{\rm t}=2.848(1)$ and $T_{\rm t}=1.4019(2)$, was later reported by Deng and Bl\"ote using cluster Monte Carlo simulations~\cite{deng2004}. This estimate was subsequently adopted in the recent study of the three-dimensional Blume-Capel antiferromagnet by Silva \emph{et al.}~\cite{silva2023}, where the zero-field phase diagram is identical to that of the ferromagnetic model. Importantly, the phase boundary remains smooth at the tricritical point, with the continuous and first-order transition lines joining continuously without a cusp~\cite{lawrie}.

\begin{figure}
\centering
\includegraphics[width=0.55\linewidth]{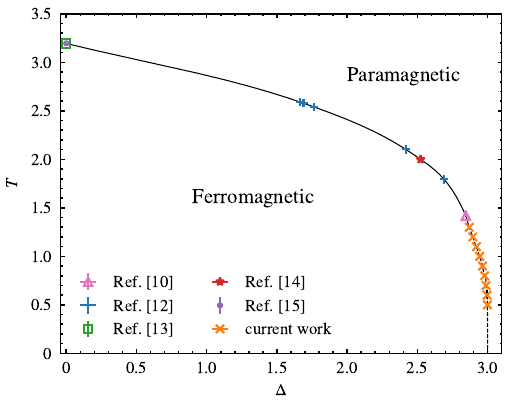}
\caption{Phase diagram of the three-dimensional spin-$1$ Blume-Capel ferromagnet in the $(\Delta,T)$ plane. The transition points investigated in the present work along the first-order segment of the phase boundary (dashed line) are shown together with previous numerical estimates from Refs.~\cite{hasenbusch10,fytas13b,zierenberg2015,moueddene24} for the continuous transition line (solid line). The triangle marks the tricritical-point estimate of Ref.~\cite{deserno97}, which serves as a reference throughout the present study.}
\label{fig:pd}
\end{figure}

Several exact and limiting results constrain the topology of the phase diagram. At zero temperature, ferromagnetic order is energetically favored whenever the interaction energy per spin, $zJ/2$, exceeds the crystal-field penalty $\Delta$, where $z$ denotes the coordination number. Consequently, the point $(\Delta_0=zJ/2,T=0)$ belongs to the phase boundary~\cite{capel}. 
The transition temperature at vanishing crystal field is not known exactly in three dimensions and must therefore be determined numerically. In the present work we focus on the simple-cubic lattice, for which $z=6$ and thus $\Delta_0=3$. For this system, one of the most accurate estimates of the tricritical point was obtained by Deserno~\cite{deserno97} using a microcanonical Monte Carlo approach, yielding $(\Delta_{\rm t},T_{\rm t}) = [2.84479(30),1.4182(55)]$. A schematic representation of the phase diagram, based on the present data together with previous numerical results from Refs.~\cite{deserno97,hasenbusch10,fytas13b,zierenberg2015,moueddene24}, is shown in Fig.~\ref{fig:pd}; see also Table~\ref{tab:delta_c}. Throughout this paper we set $J=1$ and $k_{\rm B}=1$.

Since its original formulation, the model~\eqref{eq:Hamiltonian} has attracted sustained interest and has been investigated using a broad spectrum of analytical and numerical approaches for a range of lattices, mainly in two and three dimensions, see, e.g.,
Refs.~\cite{zierenberg2015,fytas_BC,fytas2012,fytas2013}. Most work has been devoted to the two-dimensional model lattice, employing a wide range of methods, including real-space renormalization~\cite{berker1976rg}, Monte Carlo simulations, Monte Carlo renormalization-group
calculations~\cite{landau1972,kaufman1981,selke1983,selke1984,landau1986,xavier1998,deng2005,silva2006,hurt2007,malakis1,malakis2,malakis3,kwak2015},
$\epsilon$ expansions~\cite{stephen1973,chang1974,tuthill1975,wegner1975}, high- and low-temperature series expansions~\cite{fox1973,camp1975,burkhardt1976}, and 
transfer matrix calculations~\cite{beale1986,kim17,blote2019,mataragkas25b}. Despite this extensive body of work, a comprehensive finite-size scaling investigation of the first-order transition regime extending from the tricritical region down to low temperatures remains comparatively scarce in three dimensions.

\begin{table}[tb!]
\caption{Summary of transition and critical fields ($\Delta_{\infty}^{\ast}$ or $\Delta_{\rm c}$) for the spin-$1$ three-dimensional Blume-Capel ferromagnet. Results obtained in the present work from two-parameter Wang-Landau (WL) and multicanonical (MUCA) simulations at fixed temperatures are compared with literature estimates of the critical temperature $T_{\rm c}$ obtained from various Monte Carlo (MC) schemes at fixed values of the scaling field $\beta \Delta$~\cite{hasenbusch10,fytas13b,zierenberg2015,moueddene24}. Most of the literature values listed in the fourth column are taken from Ref.~\cite{hasenbusch10}, while for $\Delta = 0$ the quoted value of $T_{\rm c}$ corresponds to the average of the estimates reported in Refs.~\cite{fytas13b,moueddene24}. The last column indicates the order of the transition at the corresponding parameter set. The tricritical point separating the continuous and first-order regimes~\cite{deserno97} is included for completeness.}
\label{tab:delta_c}
        \begin{tabular}{lcclcc}
        \hline  \hline 
            $T$ & $\Delta_\infty^{\ast}$ (WL) & $\Delta^{\ast}_\infty$ or $\Delta_{\rm c}$ (MUCA) & $T_{\rm c}$ (MC)~\cite{hasenbusch10} &  $\beta\Delta$ & Order of transition\\ [0.2em]
            \hline 
            0.5 & 2.998709(2) &  &  & & First\\
            0.6 &  2.99568(1) &  & & & First\\ 
            0.7 & 2.98939(1) &  & & & First\\ 
            0.8 & 2.97886(1) &   & & & First\\
            0.9 & 2.96354(1) &  & & & First\\
            1 & 2.94362(1)  & &  & & First\\
            1.1 & 2.92004(3)  & 2.92006(2)&  & & First\\
            1.2 & 2.8945(1)  & &  & & First\\
            1.3 & 2.8695(3)  & &  & & First\\  \hline
            1.4182(55) & 2.84479(30)~\cite{deserno97}&  & & & Tricritical point\\ \hline
              &  &  & 1.793624(3)  & 1.5  & Second\\
            2 &  & 2.523(6)~\cite{zierenberg2015} &  &  & Second\\
              &  &  & 2.1025586(9) & 1.15  & Second\\
              &  &  & 2.5417974(5) & $\ln{2}$  & Second\\
              &  &  & 2.5791693(2) & 0.655  & Second\\
              &  &  & 2.5928821(3) & 0.641  & Second\\ 
              &  &  & 3.1962(3)~\cite{fytas13b,moueddene24}    & 0                  & Second \\ 
             \hline \hline
        \end{tabular}
\end{table}

To address this issue, we undertake a systematic investigation of the three-dimensional Blume-Capel model on the simple-cubic lattice, focusing on the first-order segment of the phase boundary and its connection to the tricritical region. Our study combines multicanonical simulations with two-parameter Wang-Landau sampling, allowing for an accurate characterization of the thermodynamic behavior over a broad range of temperatures and crystal-field couplings. This generalized-ensemble framework provides direct access to the coexistence properties of the system and enables a detailed examination of the finite-size scaling behavior along the transition line. Particular attention is devoted to the evolution of the energy probability density function, the development of phase coexistence, and the scaling properties of the free-energy barrier and the interfacial tension. By analyzing a sequence of transition points extending from the vicinity of the tricritical point to deep within the first-order regime, we elucidate the gradual crossover from weak to strong first-order behavior and establish a comprehensive picture of the transition characteristics throughout this region of the phase diagram.

The remainder of this paper is organized as follows. In Section~\ref{sec:muca}, we present the multicanonical simulation framework employed in this work. After introducing the relevant thermodynamic observables, we discuss the finite-size scaling methodology used to characterize the first-order transition regime and to analyze quantities associated with phase coexistence and interfacial phenomena. Section~\ref{sec:WL} is devoted to the two-parameter Wang-Landau approach. We provide details of the joint-density-of-states calculations and employ a field-mixing analysis to investigate the tricritical region and its connection to the first-order segment of the phase boundary. Finally, Section~\ref{sec:conclusions} summarizes our main findings and discusses their implications for the understanding of first-order phase transitions and tricriticality in the three-dimensional Blume-Capel model.

\section{Multicanonical simulations}
\label{sec:muca}

The first-order transition regime of the three-dimensional Blume-Capel model is investigated using multicanonical simulations. Owing to their ability to overcome free-energy barriers and efficiently sample phase-coexistence configurations, multicanonical methods are particularly well suited for the study of discontinuous phase transitions. They provide direct access to the probability distributions and finite-size scaling properties that characterize first-order behavior. In the following, we describe the implementation of the multicanonical algorithm, introduce the observables considered in the analysis, and present the finite-size scaling framework used throughout this work.

\subsection{Method and observables}
\label{sec:method_obs}

The multicanonical (MUCA) method~\cite{berg92} replaces the Boltzmann factor $e^{-\beta E}$ ($\beta \equiv 1/T$) by suitably chosen generalized weights that promote a nearly uniform sampling of the relevant energy range. This strategy is particularly effective for first-order phase transitions, where canonical simulations suffer from exponentially suppressed tunneling events between coexisting phases~\cite{janke03,gross18}. For the Blume-Capel model, whose density of states depends on the two energy variables $E_J$ and $E_\Delta$, we apply the multicanonical procedure to the crystal-field contribution $E_\Delta$ while keeping the temperature fixed. This permits continuous reweighting in the crystal-field parameter $\Delta$.

Starting from the partition function of the Blume-Capel model at zero magnetic field 
\begin{equation}
\label{eq:Z}
    \mathcal{Z}(\beta, \mu) = \sum_{E_J, E_\Delta} \Gamma(E_J, E_\Delta) e^{(-\beta E_J - \mu E_\Delta)}, 
\end{equation}
where $\mu \equiv \Delta/T$, the multicanonical partition function reads
\begin{equation}
    {\cal Z}_\mathrm{MUCA}=\sum_{E_J, E_\Delta} \Gamma(E_J, E_\Delta) e^{-\beta E_J} W(E_\Delta),
\label{eq:zng}
\end{equation}
where the generalized weight $W(E_\Delta)$ replaces the Boltzmann factor associated with the crystal-field energy. A flat marginal distribution in $E_\Delta$ is obtained for
\begin{equation}
    W(E_{\Delta}) \propto {\cal Z}_\mathrm{MUCA} \left[ \sum_{E_J} \Gamma(E_J, E_\Delta)  e^{-\beta E_J}  \right]^{-1}.
\label{eq:wmuca}
\end{equation}

In order to iteratively approximate the generalized weights $W\left(E_\Delta\right)$, we sampled histograms of the crystal-field energy. Supposing that at the $n^\text{th}$ iteration, a histogram $H^{(n)}(E_\Delta)$ was sampled, then its average should depend on the weight of the iteration $W^{(n)}\left(E_\Delta\right)$ as
\begin{equation}
    \langle H^{(n)}(E_\Delta)\rangle \propto \sum_{E_J} \Gamma(E_J, E_\Delta) e^{-\beta E_{J}} W^{(n)}(E_\Delta).
\label{eq:hmuca}
\end{equation}
From Eqs.~\eqref{eq:wmuca} and \eqref{eq:hmuca}, it follows that $\langle H^{(n)}(E_\Delta)\rangle\propto W^{(n)}(E_\Delta) / W(E_\Delta)$. This justifies our weight modification scheme $W^{(n+1)}\left(E_\Delta\right) = W^{(n)}\left(E_\Delta\right)/H^{(n)}(E_\Delta)$ for iterations to approximate $W(E_\Delta)$ producing a flat histogram.
The iterations stop when the histogram satisfies a suitable flatness criterion. We tested the flatness of the histogram using the Kullback-Leibler divergence~\cite{kullback51, gross18}, which is adequate for our distributions that do not have a long tail of near-zero values. The weights are fixed after the iterations and then used for the production runs.

The multicanonical algorithm is particularly well suited for massively parallel implementations~\cite{gross18,zierenberg13}. Independent replicas evolve simultaneously using identical weights but distinct random-number sequences, while only the accumulated histograms need to be communicated between iterations. This approach has previously been applied successfully to several spin models, including the Blume-Capel and Baxter-Wu systems~\cite{mataragkas25a,zierenberg2015,zierenberg17,fytas18,fytas22,vasilopoulos22,macedo23}. The present simulations were carried out on Nvidia RTX2080 and GTX1080Ti graphics processors, allowing us to evolve up to $57,344$ independent system replicas.

For a generic observable $O=O({\sigma})$, canonical expectation values at arbitrary crystal fields are obtained through reweighting,
\begin{equation}
	\langle O\rangle_\Delta
	=
	\frac{ \left\langle O(\{\sigma\})\,e^{-\beta\Delta E_{\Delta}(\{\sigma\})} W^{-1}\left(E_{\Delta}\right) \right\rangle_\mathrm{MUCA}}
	{ \left\langle e^{-\beta\Delta E_{\Delta}(\{\sigma\})} W^{-1}\left(E_{\Delta}\right) \right\rangle_\mathrm{MUCA} }.
	\label{eq:muca-reweight}
\end{equation}
where the averages are taken in the multicanonical ensemble. To ensure numerical stability during the reweighting procedure, all exponential sums were evaluated using arbitrary-precision arithmetic provided by the {\tt GNU MPFR} library~\cite{MPFR}. Maxima of reweighted observables were located with the Brent algorithm from the GNU Scientific Library.

The simulations were performed on simple-cubic lattices with periodic boundary conditions and linear sizes in the range $8 \le L \le 34$. To investigate the evolution of the first-order transition regime, we considered temperatures from the tricritical point $T_{\rm t} = 1.4182$ down to $T=1.1$, where the transition is strongly discontinuous. We monitored the total energy,
$E = E_J + \Delta E_\Delta$,
and the magnetization $m$, from which the magnetic susceptibility
\begin{equation}\label{eq:susceptibility}
\chi = \beta N \left( \langle m^{2}\rangle - \langle m\rangle^{2} \right),
\end{equation}
and the specific heat
\begin{equation}\label{eq:specific-heat}
C = \beta^2 N \left( \langle E^2 \rangle - \langle E \rangle^2 \right)
\end{equation}
were obtained ($N = L^{3}$ denotes the total number of spins).
It is worth mentioning that an additional characterization of the transition can be obtained from the fluctuations of the density of nonzero spins, $e_\Delta=E_\Delta/N$, as discussed in detail by Silva \emph{et al.}~\cite{silva2023}. Defining the corresponding susceptibility as
$\chi_\Delta=\beta N\left(\langle e_\Delta^2\rangle-\langle e_\Delta\rangle^2\right)$,
its leading finite-size scaling behavior is expected to be identical to that of the specific heat, since both are susceptibilities associated with energy-like operators conjugate to the crystal field $\Delta$.

As our primary focus is the characterization of the first-order transition regime, we analyzed the probability density function of the crystal-field energy per spin,
$P(e_\Delta)$, with $e_\Delta = E_\Delta/N$,
following the approach of Ref.~\cite{mataragkas25a}. For finite systems undergoing a first-order phase transition, the emergence of a double-peaked structure signals phase coexistence and anticipates the two $\delta$-function singularities expected in the thermodynamic limit~\cite{binder87,binder84}.

\begin{figure}
    \centering
    \includegraphics[width=1.0\linewidth]{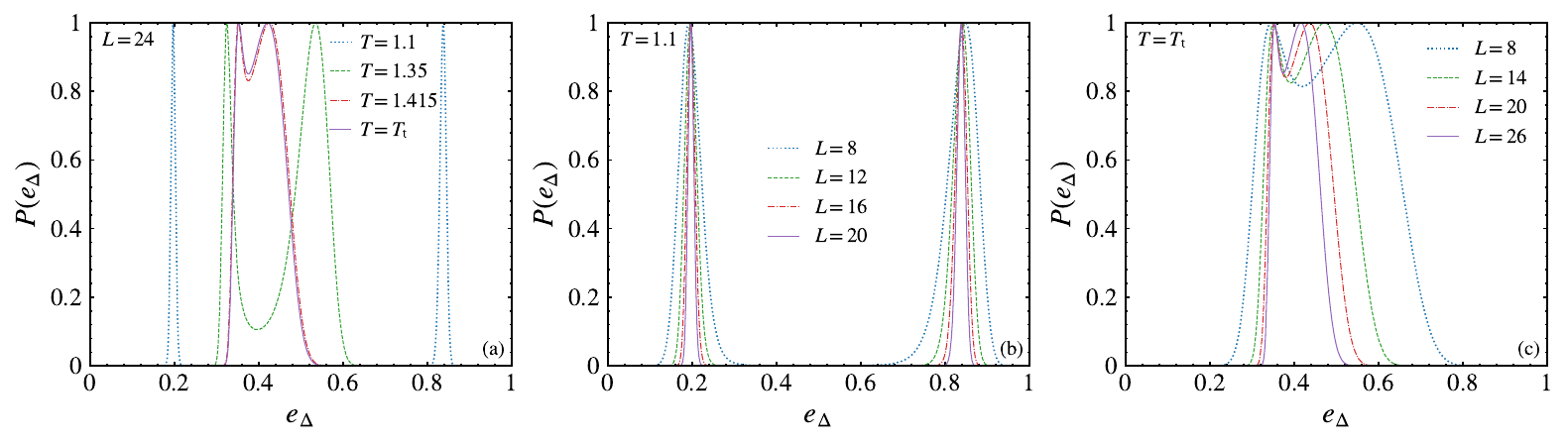}
    \caption{Normalized probability density functions $P(e_{\Delta})$ obtained from multicanonical simulations. Panel (a) shows results for a system with linear size $L=24$ at selected temperatures along the first-order transition line. Panels (b) and (c) illustrate the evolution of the distributions with system size at $T=1.1$ in the first-order regime and at the reference tricritical temperature $T=T_{\rm t}=1.4182$, respectively.}
    \label{fig:distributions}
\end{figure}

Representative normalized distributions $P(e_\Delta)$ for several temperatures and system sizes are shown in Fig.~\ref{fig:distributions}. Deep inside the first-order regime, the minimum between the two peaks becomes increasingly suppressed as the temperature decreases. A similarly pronounced effect is observed with increasing system size [Fig.~\ref{fig:distributions}(b)], reflecting the progressive suppression of intermediate configurations between the coexisting phases. In contrast, near the proposed tricritical point [Fig.~\ref{fig:distributions}(c)], the distributions retain a double-peaked structure, but the barrier separating the peaks is considerably reduced and exhibits only a weak dependence on system size. The finite-size scaling behavior associated with these features is analyzed in detail below.

\subsection{Finite-size scaling analysis}
\label{sec:fss}

Before discussing the finite-size scaling analysis, let us briefly comment on the fitting procedure employed throughout this work. In all fits, we restricted the analysis to lattice sizes $L\geq L_{\rm min}$ and assessed the quality of the fits using the standard $\chi^2$ goodness-of-fit criterion. Specifically, a fit was considered acceptable only if the corresponding quality-of-fit parameter satisfied $10\% < Q < 90\%$~\cite{press92}. The value of $L_{\rm min}$ was chosen individually for each analysis so as to minimize the influence of subleading finite-size corrections while retaining a statistically meaningful number of data points.

\begin{figure}
    \centering
    \includegraphics[width=0.55\linewidth]{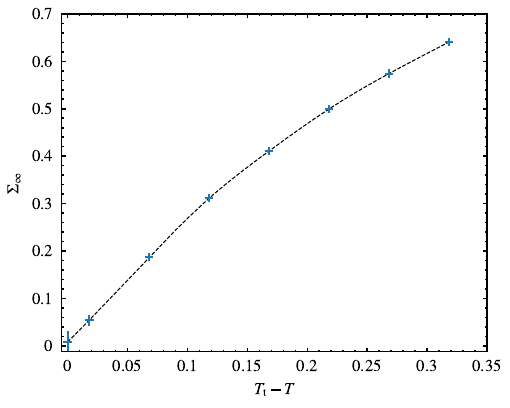}
    \caption{Thermodynamic-limit surface tension $\Sigma_{\infty}$ as a function of the temperature distance from the tricritical point, $T_{\rm t}-T$. The values of $\Sigma_{\infty}$ are obtained from fits of the finite-size data to Eq.~\eqref{eq:tension}. The dashed line is a quadratic polynomial fit, illustrating the monotonic approach of the surface tension to zero near tricriticality.}
    \label{fig:tension}
\end{figure}

The observations in Fig.~\ref{fig:distributions} naturally lead us to investigate the systematic behavior of the surface tension that may characterize the transition as suggested by Lee and Kosterlitz~\cite{lee90,lee91}. The multicanonical method is instrumental for this purpose since it allows the direct estimation of the barrier associated with suppressing intermediate states during a first-order phase transition. Although the equal-weight prescription generally provides more accurate finite-size estimates of transition points~\cite{borgs92}, the equal-height criterion is the natural choice for defining the Lee-Kosterlitz free-energy barrier and the associated interface tension. Accordingly, considering distributions with two peaks of equal height (eqh)~\cite{borgs92}, such as the ones shown in Fig.~\ref{fig:distributions}, allows one to extract the free-energy like barrier in the $e_{\Delta}$-space,
\begin{equation}
	\Delta F_{L} = \frac{1}{2\beta\Delta}\ln{\left(\frac{P_{\rm
			max}}{P_{\rm min}}\right)_{\rm eqh}},
\end{equation}
where $P_{\rm max}$ and $P_{\rm min}$ are the maximum and local minimum of the distribution $P(e_{\Delta})$, respectively. We note that the twofold degeneracy of the ferromagnetic ordered phase does not influence the present analysis, since both symmetry-related ordered states have identical crystal-field energies and therefore contribute equally to the ordered-phase peak of $P(e_{\Delta})$. Consequently, the free-energy barrier and the corresponding interface tension are unaffected by this degeneracy. The resulting barrier connects a spin-$0$ dominated regime ($e_{\Delta}$ small) and a spin-$\pm1$ rich phase ($e_{\Delta}$ large). The corresponding surface tension $\Sigma_{L} = \Delta F_{L}/L^2$ is expected to exhibit the finite-size scaling behavior in three dimensions
\begin{equation}
\label{eq:tension}
\Sigma_{L}  = \Sigma_{\infty} + \mathcal{A}_{\Sigma}L^{-2} + \mathcal{O}\left(L^{-4}\right), 
\end{equation}
where $\Sigma_{\infty}$ denotes the thermodynamic-limit surface tension and the $\mathcal{A}$ parameters introduced throughout this work represent non-universal amplitudes. Higher-order corrections may also be present~\cite{Nussbaumer2006,Nussbaumer2008,Bittner2009}. The resulting estimates of $\Sigma_{\infty}$ are plotted as a function of temperature in Fig.~\ref{fig:tension}. As expected, the surface tension increases monotonically upon moving deeper into the first-order regime, i.e., as the temperature decreases below the tricritical point. This behavior is consistent with recent findings for the two-dimensional Blume-Capel model~\cite{mataragkas25a,kim17,mataragkas25b}. Conversely, upon approaching the tricritical temperature from the first-order side, the surface tension decreases and tends to zero, reflecting the disappearance of phase coexistence at tricriticality. The approximately linear approach of $\Sigma_{\infty}$ to zero in the vicinity of the tricritical point is also consistent with theoretical predictions for tricritical systems~\cite{schick86,cirillo96}. Indeed, at the reference tricritical temperature $T_{\rm t}=1.4182$~\cite{deserno97}, we obtain $\Sigma_{\infty} = 0.008 \pm 0.020$, a value fully compatible with a vanishing surface tension within statistical uncertainty.

\begin{figure}
    \centering
    \includegraphics[width=1.0\linewidth]{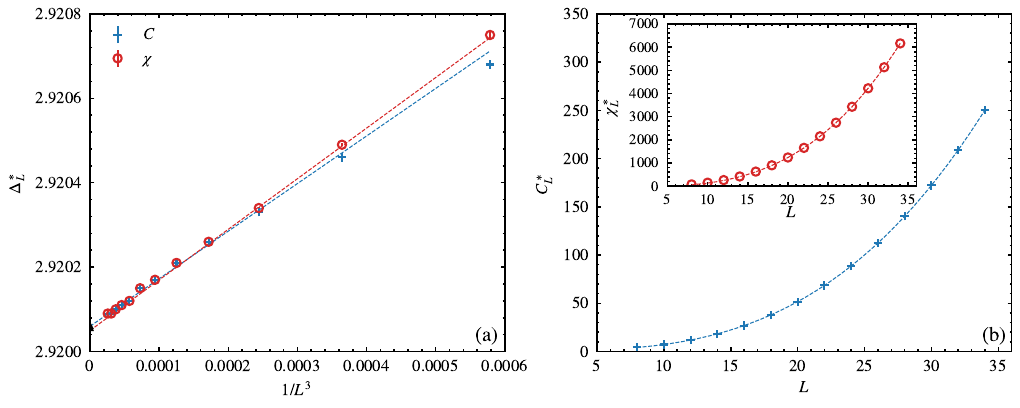}
    \caption{(a) Finite-size shift of the pseudo-transition fields $\Delta_L^\ast$ obtained from the locations of the specific-heat ($C$) and magnetic-susceptibility ($\chi$) maxima. The dashed lines represent fits to Eq.~\eqref{eq:shift} using data with $L \geq L_{\rm min} = 10$. (b) Volume scaling of the specific-heat maxima (main panel) and magnetic-susceptibility maxima (inset) at $T = 1.1$. The dashed lines represent fits to the first-order finite-size scaling forms of Eq.~\eqref{eq:scaling}, using data with $L \geq L_{\rm min} = 12$.}
    \label{fig:fss_scaling}
\end{figure}

We now focus our analysis deep in the first-order transition regime at $T = 1.1$. We first examine the system-size scaling behavior of the pseudo-transition fields $\Delta_{L}^{\ast}$ obtained from the maxima of the specific heat and magnetic susceptibility. Figure~\ref{fig:fss_scaling}(a) presents the shift behavior of $\Delta_{L}^{\ast}$ and our data indicate an excellent fit to the expected scaling form 
\begin{equation}
\label{eq:shift}
\Delta_{L}^{\ast} = \Delta_{\infty}^{\ast}+\mathcal{A}_{\Delta}L^{-d} (1+\mathcal{O}(L^{-d})),
\end{equation}
where $d = 3$, and $\Delta^{\ast}_\infty = 2.92006(2)$ is the desired transition field from both fits that compares excellently with the value $2.92004(3)$ obtained from the field-mixing analysis below; see also Table~\ref{tab:delta_c}. Figure~\ref{fig:fss_scaling}(b) explores further aspects of the characteristic scaling associated with first-order phase transitions. In the specific heat and magnetic susceptibility, we expect that their maxima along the $\Delta$ axis behave as $C^{\ast}_L \sim L^{d}$ and $\chi^{\ast}_L \sim L^{d}$ in the leading order of $L$, respectively. For systems with periodic boundary conditions, corrections to first-order scaling are expected to occur in inverse powers of the volume~\cite{janke1,janke2}. In three dimensions, this leads to
\begin{equation}
    C^{\ast}_L = \mathcal{A}_C L^{x} (1+\mathcal{O}(L^{-d})) \;\; ; \;\;    \chi^{\ast}_L = \mathcal{A}_\chi L^{x} (1+\mathcal{O}(L^{-d})). \label{eq:scaling} 
\end{equation}
The scaling Ansätze of Eq.~\eqref{eq:scaling} provides excellent descriptions of the numerical data, as illustrated in Fig.~\ref{fig:fss_scaling}(b). Fitting the specific-heat and magnetic-susceptibility maxima yields $x = 3$ for both observables, with an accuracy of approximately $10^{-4}$. These estimates are in excellent agreement with the theoretical prediction $x=d=3$, providing further evidence for the first-order nature of the transition. Since the fitted exponent is essentially equal to three, the leading correction term proportional to $L^{-3}$ gives rise asymptotically to a constant background contribution in addition to the dominant volume scaling.

\section{Two-parameter Wang-Landau simulations}
\label{sec:WL}

In this section, we employ the field-mixing approach in conjunction with the Wang-Landau joint density of states to identify transition points along the first-order segment of the phase boundary. We first describe the numerical implementation of the two-parameter Wang-Landau simulations and then outline the procedure used to locate finite-size pseudo-transition points and extrapolate them to the thermodynamic limit. 

\subsection{Joint-density-of-states calculations}
\label{sec:sim_details}

We employ the standard Wang-Landau algorithm~\cite{WL1,WL2} to estimate the joint density of states $\Gamma(E_J,E_\Delta)$ of the three-dimensional Blume-Capel model on simple-cubic lattices of size $L\times L\times L$ with periodic boundary conditions. The implementation follows closely the procedures used previously for the square- and triangular-lattice Blume-Capel models~\cite{mataragkas25a,kwak2015}. However, the substantially larger number of accessible states in three dimensions necessitates a relaxation of the convergence criteria employed in those studies.
To account for the rapid growth of the $(E_J,E_\Delta)$ state space with increasing system size, the histogram flatness criterion was set to $95\%$ for $L=6$, $90\%$ for $L=8$, and $80\%$ for $L \geq 10$. For illustration, the largest lattice considered here ($L=16$) contains $48,508,757$ accessible states in the $(E_J,E_\Delta)$ plane, compared with $10,424,521$ states for the largest square-lattice calculation reported previously ($L=48$)~\cite{mataragkas25a}. The final modification factor was chosen as $f_{\rm min} = 10^{-7}$ for $L\leq12$ and $f_{\rm min}=10^{-6}$ for $L=16$. Joint densities of states were obtained for system sizes $L\in \{6,8,10,12,16\}$.

\begin{figure}
    \centering
    \includegraphics[width=1.0\linewidth]{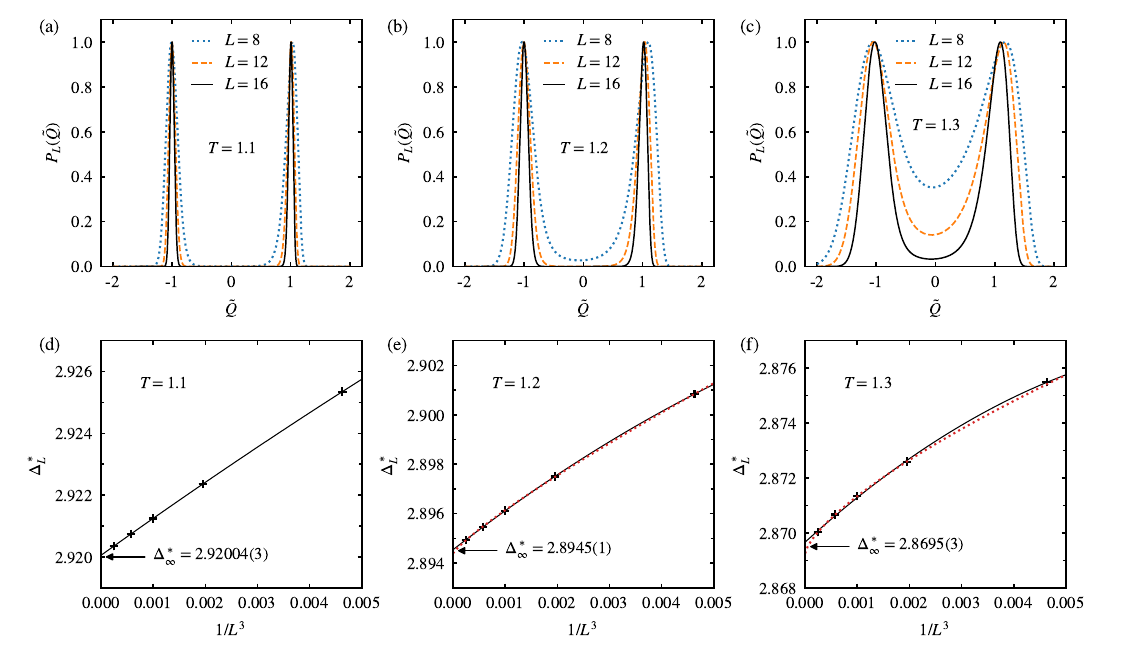}
    \caption{Field-mixing analysis of the first-order transition regime. Panels (a)--(c) show the probability distribution $P_L(\tilde{\mathcal{Q}})$ of the rescaled field-mixing variable $\tilde{\mathcal{Q}}$ at the finite-size pseudo-transition point $\Delta_L^\ast$ for $T=1.1$, $1.2$, and $1.3$, respectively. As in Fig.~\ref{fig:distributions}, the distributions have been normalized to the same peak height. Panels (d)--(f) illustrate the extrapolation of the pseudo-transition fields $\Delta_L^\ast$ to the thermodynamic limit using the expected $L^{-3}$ scaling form with corrections (solid lines) and a simple power-law fit (dotted lines), yielding the corresponding transition fields $\Delta_\infty^\ast$.}
    \label{fig:field_mixing}
\end{figure}

\subsection{First-order transition points}
\label{sec:field-mixing}

The principal outcome of the Wang-Landau calculations is the determination of transition points along the first-order segment of the phase boundary. Since the partition function of the Blume-Capel model depends on both the inverse temperature $\beta$ and crystal field $\mu$, the coexistence curve is generally not aligned with either parameter axis. Consequently, finite-size analyses performed solely along the $\beta$ or $\mu$ directions may be affected by asymmetry effects, particularly in the vicinity of the tricritical point. To address this issue, we examine the field-mixing approach originally developed for fluid and magnetic systems~\cite{wilding92,bruce92,wilding96,plascak13}. This framework has previously been employed with considerable success in studies of the two-dimensional Blume-Capel model~\cite{mataragkas25a,silva2006,kwak2015}, and here we investigate its performance in the three-dimensional case.

Here we examine the applicability of the field-mixing method to the three-dimensional model. Since the theoretical framework has been reviewed extensively elsewhere~\cite{mataragkas25a,kwak2015,wilding92,bruce92,wilding96,plascak13}, we restrict our discussion to the scaling variable conjugate to the field across the phase boundary, which plays a central role in locating the coexistence curve. It is defined as
\begin{equation}
    \mathcal{Q} = \frac{e_\Delta - s e_J}{1-rs},
\end{equation}
where $e_J \equiv E_J/N$, while $r$ and $s$ are the mixing parameters associated with the scaling fields across and tangent to the phase boundary. These non-universal parameters generally depend on system size and are determined numerically by requiring the probability distribution $P(\mathcal{Q})$ to exhibit a symmetric double-peaked structure at the pseudo-transition point. Following Refs.~\cite{mataragkas25a,kwak2015}, we impose both equal-population and equal-height conditions on the two peaks of the distribution. Since the parameter $r$ only rescales $\mathcal{Q}$ and does not affect the shape of $P(\mathcal{Q})$, only the parameter $s$ needs to be determined explicitly.

For the analysis of the distribution shape, we employ the normalized variable
\begin{equation}
\tilde{\mathcal{Q}} =
\frac{\mathcal{Q}-\bar{\mathcal{Q}}}{\sigma_{\mathcal{Q}}},
\end{equation}
where $\bar{\mathcal{Q}}$ and $\sigma_{\mathcal{Q}}$ denote the mean and standard deviation of $\mathcal{Q}$, respectively. A small Gaussian broadening was introduced to estimate peak heights from the discrete distributions, using widths of $0.01$ for $L\leq8$ and $0.005$ for $L\geq 10$.

For a given temperature $T$ and system size $L$, the mixing parameter $s_L$ and pseudo-transition field $\Delta_L^\ast$ were determined to satisfy the equal-population and equal-height conditions simultaneously. In practice, we first obtained $\Delta^\ast(s)$ from the equal-population condition and then iteratively adjusted $s$ until the equal-height criterion was also satisfied. Using this field-mixing procedure, we determined pseudo-transition fields for temperatures up to $T=1.3$ using the Wang-Landau joint densities of states available for system sizes up to $L=16$. In the deep first-order regime at low temperatures, the mixing parameter is small, typically of the order of $s\sim10^{-2}$, so that the scaling variable $\mathcal{Q}$ is dominated by the nonzero-spin density $e_\Delta$, as expected from the nearly vertical phase boundary. As the temperature increases, however, the mixing parameter grows steadily, reaching values of approximately $s\simeq0.2$ at $T=1.3$. This behavior indicates that, upon approaching the tricritical region, the ordering scaling field becomes an increasingly strong linear combination of $e_\Delta$ and $e_J$, reflecting the gradual change in the orientation of the phase boundary.

Figure~\ref{fig:field_mixing} presents representative distributions $P(\tilde{\mathcal{Q}})$ at the corresponding pairs $(s_L,\Delta_L^\ast)$ for $T=1.1$, $1.2$, and $1.3$ (for the full set of results, we refer the reader to Table~\ref{tab:delta_c}). The distributions display the expected double-peaked structure, while the probability of intermediate states decreases systematically with increasing system size. This behavior is characteristic of phase coexistence and corroborates the first-order nature of the transition, as already illustrated in Figs.~\ref{fig:distributions} and~\ref{fig:tension}. The associated pseudo-transition fields exhibit increasingly strong finite-size effects upon approaching the tricritical region, necessitating a careful extrapolation to the thermodynamic limit.

\begin{figure}
    \centering
    \includegraphics[width=0.48\linewidth]{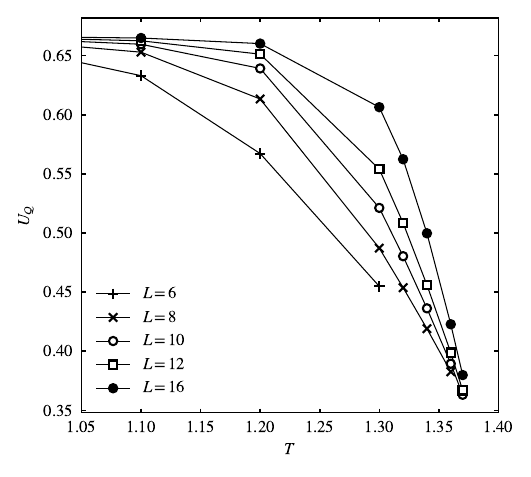}
    \caption{Temperature dependence of the fourth-order cumulant $U_{\mathcal Q}$ of the normalized scaling variable $\tilde{\mathcal Q}$. The sharp decrease of $U_{\mathcal Q}$ upon approaching the tricritical region signals the progressive suppression of the double-peaked structure in $P(\tilde{\mathcal Q})$ and coincides with the temperature range where the field-mixing analysis ceases to satisfy simultaneously the equal-population and equal-height conditions.}
    %\caption{Fourth-order cumulant $U_\mathcal{Q}$ of the scaling variable $\tilde{\mathcal{Q}}$ as a function of temperature $T$.}
    \label{fig:binder}
\end{figure}

At $T=1.1$, the pseudo-transition fields $\Delta_L^\ast$ obtained from the field-mixing analysis follow closely the expected $L^{-3}$ finite-size scaling behavior of a first-order transition, as illustrated by the solid line in Fig.~\ref{fig:field_mixing}(d). Extrapolation to the thermodynamic limit yields $\Delta_\infty^\ast=2.92004(3)$, in excellent agreement with the estimate $\Delta_\infty^\ast=2.92006(2)$ obtained independently from the multicanonical analysis of the specific-heat and magnetic-susceptibility maxima shown in Fig.~\ref{fig:fss_scaling}. At the higher temperature $T=1.2$, finite-size effects become noticeably stronger and deviations from the leading $L^{-3}$ behavior are more pronounced. As shown in Fig.~\ref{fig:field_mixing}(e), the data were analyzed using two different Ansätze. The solid line corresponds to the first-order scaling form of Eq.~\eqref{eq:shift}, which requires a sizable $L^{-6}$ correction term. The dotted line represents a more general power-law fit, $\Delta_L^\ast=\Delta_\infty^\ast+aL^{-x}$, yielding an effective exponent $x \approx 2.557$. The deviation of this exponent from the asymptotic value $x=3$ further reflects the importance of finite-size corrections for the relatively small system sizes accessible to the Wang-Landau calculations. At $T=1.3$, the finite-size effects become even more pronounced. In this case, the power-law fit yields an effective exponent $x \approx 2.120$, and a larger discrepancy is observed between the power-law extrapolation and the fit based on the expected $L^{-3}$ scaling behavior including 
$L^{-6}$ corrections.     

Increasing the temperature up to $T=1.4$, we were unable to identify pairs $(s_L,\Delta_L^\ast)$ that simultaneously satisfy the equal-population and equal-height conditions within the range of system sizes currently accessible to our Wang--Landau calculations. The highest temperature for which a solution can be obtained depends on the lattice size, but for all systems considered the procedure ceases to converge in the interval $1.3 < T < 1.4$. To quantify this trend, we consider the fourth-order cumulant
\begin{equation}
U_{\mathcal Q}
=
1-\frac{\langle \tilde{\mathcal Q}^{\,4}\rangle}
{3\langle \tilde{\mathcal Q}^{\,2}\rangle^2}.
\end{equation}
As shown in Fig.~\ref{fig:binder}, $U_{\mathcal Q}$ decreases rapidly as the tricritical region is approached. Empirically, the equal-height and equal-population conditions become increasingly difficult to satisfy once $U_{\mathcal Q}$ falls below approximately $0.3$. In this regime, the probability distribution $P(\tilde{\mathcal Q})$ develops only a shallow minimum between its two peaks, leading to numerical instabilities in the solution of the coupled equations for $(s_L,\Delta_L^\ast)$. Consequently, for the discrete Wang-Landau data sets available up to $L = 16$, we were unable to extend the analysis closer to the tricritical region.

This limitation nevertheless allows us to examine the behavior of $U_{\mathcal Q}$ over the temperature range where solutions for $(s_L,\Delta_L^\ast)$ are available. Figure~\ref{fig:binder} shows the cumulant along the coexistence curve for $T < 1.4$. Within this range, no common crossing point is observed among different system sizes. This suggests that the tricritical temperature lies above the highest temperature accessible to the present analysis, consistent with previous estimates placing it near $T_{\rm t}\approx 1.42$~\cite{deserno97}. However, the available data do not allow for an independent determination of the tricritical point. It is nevertheless interesting to note that the observed decrease of $U_{\mathcal{q}}$ appears qualitatively consistent with the expected tricritical behavior reported by Deng and Blöte~\cite{deng2004} and by Silva \emph{et al.}~\cite{silva2023}, although the present data do not extend sufficiently close to the tricritical point to permit a quantitative comparison or an extrapolation to the expected limiting value.

Closing we would like to note an interesting distinction from the two-dimensional Blume-Capel model~\cite{mataragkas25a}: although the field-mixing procedure remains robust throughout the first-order regime, its practical implementation becomes considerably more challenging in the vicinity of the tricritical point. Our results suggest that this is not a limitation of the field-mixing formalism itself, but rather of its numerical implementation for the presently accessible system sizes. As the two peaks of the mixed-variable distribution become only weakly separated, the equal-height and equal-population conditions lose discriminatory power, making the determination of the mixing parameter increasingly sensitive to the discretization of the joint density of states. Whether this difficulty can be overcome by substantially larger system sizes or higher-resolution Wang-Landau sampling remains an interesting topic for future investigation.

\section{Conclusions}
\label{sec:conclusions}

In summary, the combination of multicanonical and Wang-Landau methodologies has enabled a detailed investigation of first-order phase transitions in the three-dimensional Blume-Capel ferromagnet over an extended region of the phase diagram. The resulting transition points reproduce and extend the established phase boundary, while the observed scaling of response functions, probability distributions, free-energy barriers, and surface tensions consistently supports the first-order character of the transition below tricriticality. The vanishing of the thermodynamic-limit surface tension at the tricritical point further highlights its usefulness as a complementary indicator of tricritical behavior. Finally, the limited effectiveness of the field-mixing analysis in the three-dimensional case suggests that alternative approaches may be required for a precise finite-size characterization of tricriticality in higher-dimensional Blume-Capel systems. These findings contribute to a more complete understanding of the interplay between tricriticality and first-order phase transitions in one of the paradigmatic models of statistical physics.

\section*{Acknowledgments}
Part of the numerical calculations reported in this paper were performed at the High-Performance Computing cluster CERES of the University of Essex. The work of A. Vasilopoulos and N.G. Fytas was supported by the Engineering and Physical Sciences Research Council (Grant No. EP/X026116/1). The work of Y. Kim and D.-H. Kim was supported by the Regional Innovation System \& Education (RISE) program through the Gwangju RISE Center, funded by the Ministry of Education and the Gwangju Metropolitan City, Republic of Korea (2026-RISE-05-001).
\printcredits

\bibliographystyle{unsrt}

\bibliography{cas-refs}

@article{blume,
  title={{Theory of the first-order magnetic phase change in $UO_2$}},
  author={Blume, M},
  journal={Phys. Rev.},
  volume={141},
  number={2},
  pages={517},
  year={1966},
  publisher={APS}
}

@article{capel,
  title={{On the possibility transitions of first-order in Ising systems with zero-field phase of triplet ions splitting}},
  author={Capel, H},
  journal={Physica (Utr.)},
  volume={32},
  number={966},
  pages={10--1016},
  year={1966}
}

@book{lawrie,
  title={{Phase Transitions and Critical Phenomena vol 9, C. Domb, J.L.
Lebowitz}},
  author={Lawrie, I D and Sarbach, S},
  year={1984},
  publisher={London: Academic) p}
}

@article{deng2004,
  title = {{Constrained tricritical Blume-Capel model in three dimensions}},
  author = {Deng, Youjin and Bl\"ote, Henk W. J.},
  journal = {Phys. Rev. E},
  volume = {70},
  issue = {4},
  pages = {046111},
  numpages = {11},
  year = {2004},
  month = {Oct},
  publisher = {American Physical Society},
  doi = {10.1103/PhysRevE.70.046111},
  url = {https://link.aps.org/doi/10.1103/PhysRevE.70.046111}
}

@article{silva2023,
  title = {{Multicritical bifurcation and first-order phase transitions in a three-dimensional Blume-Capel antiferromagnet}},
  author = {Silva, Daniel and Buend\'{\i}a, Gloria M. and Rikvold, Per Arne},
  journal = {Phys. Rev. E},
  volume = {108},
  issue = {2},
  pages = {024122},
  numpages = {13},
  year = {2023},
  month = {Aug},
  publisher = {American Physical Society},
  doi = {10.1103/PhysRevE.108.024122},
  url = {https://link.aps.org/doi/10.1103/PhysRevE.108.024122}
}

@article{schick86,
  title = {{Spin-1 model of a microemulsion}},
  author = {Schick, M. and Shih, Wei-Heng},
  journal = {Phys. Rev. B},
  volume = {34},
  issue = {3},
  pages = {1797--1801},
  numpages = {0},
  year = {1986},
  month = {Aug},
  publisher = {American Physical Society},
  doi = {10.1103/PhysRevB.34.1797},
  url = {https://link.aps.org/doi/10.1103/PhysRevB.34.1797}
}

@article{cirillo96,
	author = {Cirillo, Emilio N. M. and Olivieri, Enzo},
	date = {1996/05/01},
	doi = {10.1007/BF02183739},
	id = {Cirillo1996},
	isbn = {1572-9613},
	journal = {Journal of Statistical Physics},
	number = {3},
	pages = {473--554},
	title = {{Metastability and nucleation for the Blume-Capel model. Different mechanisms of transition}},
	url = {https://doi.org/10.1007/BF02183739},
	volume = {83},
	year = {1996}}

@article{selke-10,
  title={{Monte Carlo study of mixed-spin S=(1/2, 1) Ising ferrimagnets}},
  author={Selke, W and Oitmaa, J},
  journal={J. Phys. C},
  volume={22},
  number={7},
  pages={076004},
  year={2010},
  publisher={IOP Publishing}
}

@article{fytas13,
  title={{Wetting and interfacial adsorption in the Blume-Capel model on the square lattice}},
  author={Fytas, N G and Selke, W},
  journal={Eur. Phys. J. B},
  volume={86},
  pages={1--7},
  year={2013},
  publisher={Springer}
}

@article{leila,
doi = {10.1088/1742-5468/ad1d60},
url = {https://dx.doi.org/10.1088/1742-5468/ad1d60},
year = {{2024}},
month = {Feb},
publisher = {IOP Publishing},
volume = {2024},
number = {2},
pages = {023206},
author = {L Moueddene and N G Fytas and Y Holovatch and R Kenna and B Berche},
title = {{Critical and tricritical singularities from small-scale Monte Carlo simulations: the Blume–Capel model in two dimensions}},
journal = {J. Stat. Mech.}
}

@article{macedo24,
doi = {10.1088/1742-5468/ad784e},
url = {https://dx.doi.org/10.1088/1742-5468/ad784e},
year = {2024},
month = {oct},
publisher = {IOP Publishing},
volume = {{2024}},
number = {10},
pages = {103204},
author = {A R S Macêdo and J A Plascak and A Vasilopoulos and N G Fytas and M Akritidis and M Weigel},
title = {{Universal energy and magnetisation distributions in the Blume–Capel and Baxter–Wu models}},
journal = {J. Stat. Mech.}
}

@article{fytas_BC,
  title={{Wang-Landau study of the triangular Blume-Capel ferromagnet}},
  author={Fytas, N G},
  journal={Eur. Phys. J. B},
  volume={79},
  pages={21--28},
  year={2011},
  publisher={Springer}
}

@article{fytas2012,
  title={{Monte Carlo study of the triangular Blume-Capel model under bond randomness}},
  author={Theodorakis, P E and Fytas, N G},
  journal={Phys. Rev. E},
  volume={86},
  number={1},
  pages={011140},
  year={2012},
  publisher={APS}
}

@article{fytas2013,
  title={{Universality aspects of the 2d random-bond Ising and 3d Blume-Capel models}},
  author={Fytas, N G and Theodorakis, P E},
  journal={Eur. Phys. J. B},
  volume={86},
  pages={1--10},
  year={2013},
  publisher={Springer}
}

@article{zierenberg2015,
  title={{Parallel multicanonical study of the three-dimensional Blume-Capel model}},
  author={Zierenberg, J and Fytas, N G and Janke, W},
  journal={Phys. Rev. E},
  volume={91},
  number={3},
  pages={032126},
  year={2015},
  publisher={APS}
}

@article{berker1976rg,
  title={{Blume-Emery-Griffiths-Potts model in two dimensions: Phase diagram and critical properties from a position-space renormalization group}},
  author={Berker, A Nihat and Wortis, M},
  journal={Phys. Rev. B},
  volume={14},
  number={11},
  pages={4946},
  year={1976},
  publisher={APS}
}

@article{landau1972,
  title={{Magnetic tricritical points in Ising antiferromagnets}},
  author={Landau, D P},
  journal={Phys. Rev. Lett.},
  volume={28},
  number={7},
  pages={449},
  year={1972},
  publisher={APS}
}

@article{kaufman1981,
  title={{Three-component model and tricritical points: A renormalization-group study. Two dimensions}},
  author={Kaufman, M and Griffiths, R B and Yeomans, J M and Fisher, M E},
  journal={Phys. Rev. B},
  volume={23},
  number={7},
  pages={3448},
  year={1981},
  publisher={APS}
}

@article{selke1983,
  title={{Interface properties of the two-dimensional Blume-Emery-Griffiths model}},
  author={Selke, W and Yeomans, J},
  journal={J. Phys. A},
  volume={16},
  number={12},
  pages={2789},
  year={1983},
  publisher={IOP Publishing}
}

@article{selke1984,
  title={{Interfacial adsorption in the two-dimensional Blume-Capel model}},
  author={Selke, W and Huse, D A and Kroll, D M},
  journal={J. Phys. A},
  volume={17},
  number={15},
  pages={3019},
  year={1984},
  publisher={IOP Publishing}
}

@article{landau1986,
  title={{Monte Carlo renormalization-group study of tricritical behavior in two dimensions}},
  author={Landau, D P and Swendsen, R H},
  journal={Phys. Rev. B},
  volume={33},
  number={11},
  pages={7700},
  year={1986},
  publisher={APS}
}

@article{xavier1998,
  title={{Critical behavior of the spin-3/2 Blume-Capel model in two dimensions}},
  author={Xavier, J C and Alcaraz, F C and Lara, D Pen\~{a} and Plascak, J A},
  journal={Phys. Rev. B},
  volume={57},
  number={18},
  pages={11575},
  year={1998},
  publisher={APS}
}

@article{deng2005,
  title={{Percolation between vacancies in the two-dimensional Blume-Capel model}},
  author={Deng, Youjin and Guo, Wenan and Bl{\"o}te, Henk W J},
  journal={Phys. Rev. E},
  volume={72},
  number={1},
  pages={016101},
  year={2005},
  publisher={APS}
}

@article{silva2006,
  title={{Wang-Landau Monte Carlo simulation of the Blume-Capel model}},
  author={Silva, C J and Caparica, A A and Plascak, J A},
  journal={Phys. Rev. E},
  volume={73},
  number={3},
  pages={036702},
  year={2006},
  publisher={APS}
}

@incollection{hurt2007,
  author={Hurt, D. and Eitzel, M. and Scalettar, R T and Batrouni, G. G.},
  booktitle={Computer Simulation Studies in Condensed-Matter Physics XVI},
  volume={105},
  year={2006},
  editor={Landau, D P and Lewis, S P and Sch\"{u}ttler, H.-B.},
  publisher={Springer}
}

@article{malakis1,
  title={{Strong violation of critical phenomena universality: Wang-Landau study of the two-dimensional Blume-Capel model under bond randomness}},
  author={Malakis, A and Berker, A Nihat and Hadjiagapiou, I A and Fytas, N G},
  journal={Phys. Rev. E},
  volume={79},
  number={1},
  pages={011125},
  year={2009},
  publisher={APS}
}

@article{malakis2,
  title={{Multicritical points and crossover mediating the strong violation of universality: Wang-Landau determinations in the random-bond d= 2 Blume-Capel model}},
  author={Malakis, A and Berker, A Nihat and Hadjiagapiou, I A and Fytas, N G and Papakonstantinou, T},
  journal={Phys. Rev. E},
  volume={81},
  number={4},
  pages={041113},
  year={2010},
  publisher={APS}
}

@article{malakis3,
  title = {{Universality aspects of the $d=3$ random-bond Blume-Capel model}},
  author = {Malakis, A. and Berker, A. Nihat and Fytas, N. G. and Papakonstantinou, T.},
  journal = {Phys. Rev. E},
  volume = {85},
  issue = {6},
  pages = {061106},
  numpages = {12},
  year = {2012},
  month = {Jun},
  publisher = {American Physical Society},
  doi = {10.1103/PhysRevE.85.061106},
  url = {https://link.aps.org/doi/10.1103/PhysRevE.85.061106}
}

@article{kwak2015,
  title={{First-order phase transition and tricritical scaling behavior of the Blume-Capel model: A Wang-Landau sampling approach}},
  author={Kwak, W and Jeong, J and Lee, J and Kim, D-H},
  journal={Phys. Rev. E},
  volume={92},
  number={2},
  pages={022134},
  year={2015},
  publisher={APS}
}

@article{stephen1973,
  title={{Feynman graph expansion for tricritical exponents}},
  author={Stephen, M J and McCauley Jr, J L},
  journal={Phys. Lett. A},
  volume={44},
  number={2},
  pages={89--90},
  year={1973},
  publisher={Elsevier}
}

@article{chang1974,
  title={{Renormalization-group calculations of exponents for critical points of higher order}},
  author={Chang, T S and Tuthill, G F and Stanley, H E},
  journal={Phys. Rev. B},
  volume={9},
  number={11},
  pages={4882},
  year={1974},
  publisher={APS}
}

@article{tuthill1975,
  title={{Renormalization-group calculation of the critical-point exponent $\eta$ for a critical point of arbitrary order}},
  author={Tuthill, G F and Nicoll, J F and Stanley, H E},
  journal={Phys. Rev. B},
  volume={11},
  number={11},
  pages={4579},
  year={1975},
  publisher={APS}
}

@article{wegner1975,
  title={{Exponents for critical points of higher order}},
  author={Wegner, F J},
  journal={Phys. Lett. A},
  volume={54},
  number={1},
  pages={1--2},
  year={1975},
  publisher={Elsevier}
}

@article{fox1973,
  title={{Low temperature critical behaviour of the Ising model with spin S> 1/2}},
  author={Fox, P F and Guttmann, A J},
  journal={J. Phys. C},
  volume={6},
  number={5},
  pages={913},
  year={1973},
  publisher={IOP Publishing}
}

@article{camp1975,
  title={{High-temperature series for the susceptibility of the spin-s Ising model: analysis of confluent singularities}},
  author={Camp, W J and Van Dyke, J P},
  journal={Phys. Rev. B},
  volume={11},
  number={7},
  pages={2579},
  year={1975},
  publisher={APS}
}

@article{burkhardt1976,
  title={{Critical temperatures of the spin-s Ising model}},
  author={Burkhardt, T W and Swendsen, R H},
  journal={Phys. Rev. B},
  volume={13},
  number={7},
  pages={3071},
  year={1976},
  publisher={APS}
}

@article{beale1986,
  title={{Finite-size scaling study of the two-dimensional Blume-Capel model}},
  author={Beale, P D},
  journal={Phys. Rev. B},
  volume={33},
  number={3},
  pages={1717},
  year={1986},
  publisher={APS}
}

@article{kim17,
  title={{First-order transitions and thermodynamic properties in the 2D Blume-Capel model: the transfer-matrix method revisited}},
  author={Jung, M and Kim, D-H},
  journal={Eur. Phys. J B},
  volume={90},
  pages={1--10},
  year={2017},
  publisher={Springer}
}

@article{blote2019,
  title={{Revisiting the field-driven edge transition of the tricritical two-dimensional Blume-Capel model}},
  author={Bl{\"o}te, H W J and Deng, Y},
  journal={Phys. Rev. E},
  volume={99},
  number={6},
  pages={062133},
  year={2019},
  publisher={APS}
}

@article{WL1,
  title={{Efficient, multiple-range random walk algorithm to calculate the density of states}},
  author={Wang, F and Landau, D P},
  journal={Phys. Rev. Lett.},
  volume={86},
  number={10},
  pages={2050},
  year={2001},
  publisher={APS}
}

@article{WL2,
  title={{Determining the density of states for classical statistical models: A random walk algorithm to produce a flat histogram}},
  author={Wang, F and Landau, D P},
  journal={Phys. Rev. E},
  volume={64},
  number={5},
  pages={056101},
  year={2001},
  publisher={APS}
}

@article{wilding96,
  title={{Tricritical universality in a two-dimensional spin fluid}},
  author={Wilding, N B and Nielaba, P},
  journal={Phys. Rev. E},
  volume={53},
  number={1},
  pages={926},
  year={1996},
  publisher={APS}
}

@article{bruce92,
  title={{Scaling fields and universality of the liquid-gas critical point}},
  author={Bruce, A D and Wilding, N B},
  journal={Phys. Rev. Lett.},
  volume={68},
  number={2},
  pages={193},
  year={1992},
  publisher={APS}
}

@article{wilding92,
  title={{Density fluctuations and field mixing in the critical fluid}},
  author={Wilding, N B and Bruce, A D},
  journal={ J. Phys. Condens. Matter},
  volume={4},
  number={12},
  pages={3087},
  year={1992},
  publisher={IOP Publishing}
}

@article{plascak13,
  title={{Probability distribution function of the order parameter: Mixing fields and universality}},
  author={Plascak, J A and Martins, P H L},
  journal={Comput. Phys. Commun.},
  volume={184},
  number={2},
  pages={259--269},
  year={2013},
  publisher={Elsevier}
}

@article{binder84,
  title={{Finite-size scaling at first-order phase transitions}},
  author={Binder, K and Landau, D P},
  journal={Phys. Rev. B},
  volume={30},
  number={3},
  pages={1477},
  year={1984},
  publisher={APS}
}

@article{binder87,
  title={{Theory of first-order phase transitions}},
  author={Binder, K},
  journal={Rep. Prog. Phys.},
  volume={50},
  number={7},
  pages={783},
  year={1987},
  publisher={IOP Publishing}
}

@article{berg92,
  title={{Multicanonical ensemble: A new approach to simulate first-order phase transitions}},
  author={Berg, B A and Neuhaus, T},
  journal={Phys. Rev. Lett.},
  volume={68},
  number={1},
  pages={9},
  year={1992},
  publisher={APS}
}

@incollection{janke03,
  title={{Histograms and All That}},
  author={Janke, W},
  booktitle={Computer Simulations of Surfaces and Interfaces},
  pages={137–157},
  year={2003},
  publisher={Kluwer, Dordrecht}
}

@article{gross18,
  title={{Massively parallel multicanonical simulations}},
  author={Gross, J and Zierenberg, J and Weigel, M and Janke, W},
  journal={Comput. Phys. Commun.},
  volume={224},
  pages={387--395},
  year={2018},
  publisher={Elsevier}
}

@article{kullback51,
  title={{On information and sufficiency}},
  author={Kullback, S and Leibler, R A},
  journal={Ann. Math. Stat.},
  volume={22},
  number={1},
  pages={79--86},
  year={1951},
  publisher={JSTOR}
}

@article{zierenberg13,
  title={{Scaling properties of a parallel implementation of the multicanonical algorithm}},
  author={Zierenberg, J and Marenz, M and Janke, W},
  journal={Comput. Phys. Commun.},
  volume={184},
  number={4},
  pages={1155--1160},
  year={2013},
  publisher={Elsevier}
}

@article{zierenberg17,
  title={{Scaling and universality in the phase diagram of the 2D Blume-Capel model}},
  author={Zierenberg, J and Fytas, N G and Weigel, M and Janke, W and Malakis, A},
  journal={Eur. Phys. J. Special Topics},
  volume={226},
  number={4},
  pages={789--804},
  year={2017},
  publisher={Springer}
}

@article{fytas18,
  title={{Universality from disorder in the random-bond Blume-Capel model}},
  author={Fytas, N G and Zierenberg, Johannes and Theodorakis, P E and Weigel, M and Janke, W and Malakis, A},
  journal={Phys. Rev. E},
  volume={97},
  number={4},
  pages={040102(R)},
  year={2018},
  publisher={APS}
}

@article{fytas22,
doi = {10.1088/1742-6596/2207/1/012008},
url = {https://dx.doi.org/10.1088/1742-6596/2207/1/012008},
year = {2022},
month = {mar},
publisher = {IOP Publishing},
volume = {2207},
number = {1},
pages = {012008},
author = {N G Fytas and A Vasilopoulos and E Vatansever and A Malakis and M Weigel},
title = {{Multicanonical simulations of the 2D spin-1 Baxter-Wu model in a crystal field}},
journal = {J. Phys.: Conf. Ser.}
}

@article{vasilopoulos22,
  title={{Universality in the two-dimensional dilute Baxter-Wu model}},
  author={Vasilopoulos, A and Fytas, N G and Vatansever, E and Malakis, A and Weigel, Ma},
  journal={Phys. Rev. E},
  volume={105},
  number={5},
  pages={054143},
  year={2022},
  publisher={APS}
}

@article{macedo23,
  title={{Two-dimensional dilute Baxter-Wu model: Transition order and universality}},
  author={Mac{\^e}do, ARS and Vasilopoulos, A and Akritidis, M and Plascak, J A and Fytas, N G and Weigel, M},
  journal={Phys. Rev. E},
  volume={108},
  number={2},
  pages={024140},
  year={2023},
  publisher={APS}
}

@misc{MPFR,
    note = {{The software is publicly available at \url{https://mpfr.org}.}}
}

@book{press92,
  title={Numerical recipes in C},
  author={Press, W H},
  year={1992},
  publisher={Cambridge University Press}
}

@article{lee90,
  title={{New numerical method to study phase transitions}},
  author={Lee, J and Kosterlitz, J M},
  journal={Phys. Rev. Lett.},
  volume={65},
  number={2},
  pages={137},
  year={1990},
  publisher={APS}
}

@article{lee91,
  title={{Finite-size scaling and Monte Carlo simulations of first-order phase transitions}},
  author={Lee, J and Kosterlitz, J M},
  journal={Phys. Rev. B},
  volume={43},
  number={4},
  pages={3265},
  year={1991},
  publisher={APS}
}

@article{borgs92,
  title={{Equal weight versus equal height: a numerical study of an asymmetric first-order transition}},
  author={Borgs, C and Kappler, S},
  journal={Phys. Lett. A},
  volume={171},
  number={1-2},
  pages={37--42},
  year={1992},
  publisher={Elsevier}
}

@article{Nussbaumer2006,
  title={{Monte Carlo study of the evaporation/condensation transition of Ising droplets}},
  author={Nu{\ss}baumer, A and Bittner, E and Neuhaus, T and Janke, W},
  journal={Europhys. Lett.},
  volume={75},
  number={5},
  pages={716},
  year={2006},
  publisher={IOP Publishing}
}

@article{Nussbaumer2008,
  title={{Monte Carlo study of the droplet formation-dissolution transition on different two-dimensional lattices}},
  author={Nu{\ss}baumer, A and Bittner, E and Janke, W},
  journal={Phys. Rev. E},
  volume={77},
  number={4},
  pages={041109},
  year={2008},
  publisher={APS}
}

@article{Bittner2009,
  title={{Anisotropy of the interface tension of the three-dimensional Ising model}},
  author={Bittner, E and Nu{\ss}baumer, A and Janke, W},
  journal={Nucl. Phys. B },
  volume={820},
  number={3},
  pages={694--706},
  year={2009},
  publisher={Elsevier}
}

@incollection{janke1,
    author = {Janke, W},
    title = {{First-Order Phase Transitions}},
    booktitle = {Computer Simulations of Surfaces and Interfaces, NATO Science Series, II. Mathematics, Physics and Chemistry - Vol. 114, Proceedings of the NATO Advanced Study Institute, Albena, Bulgaria, 9--20 September 2002},
    editor={D\"unweg, B. and Landau, D.P. and Milchev, A.I.},
    pages={111--135},
    publisher = {Kluwer, Dordrecht},
    year = {2003}
}

@article{janke2,
  title={{Three-dimensional 3-state Potts model revisited with new techniques}},
  author={Janke, W and Villanova, R},
  journal={Nucl. Phys. B},
  volume={489},
  number={3},
  pages={679--696},
  year={1997},
  publisher={Elsevier}
}

@article{deserno97,
  title = {{Tricriticality and the Blume-Capel model: A Monte Carlo study within the microcanonical ensemble}},
  author = {Deserno, M},
  journal = {Phys. Rev. E},
  volume = {56},
  issue = {5},
  pages = {5204--5210},
  numpages = {0},
  year = {1997},
  month = {Nov},
  publisher = {American Physical Society},
  doi = {10.1103/PhysRevE.56.5204},
  url = {https://link.aps.org/doi/10.1103/PhysRevE.56.5204}
}

@article{mataragkas25a,
  title = {{Tricriticality and finite-size scaling in the triangular Blume-Capel ferromagnet}},
  author = {Mataragkas, D and Vasilopoulos, A and Fytas, N G and Kim, D-H},
  journal = {Phys. Rev. Res.},
  volume = {7},
  issue = {1},
  pages = {013214},
  numpages = {15},
  year = {2025},
  month = {Feb},
  publisher = {American Physical Society},
  doi = {10.1103/PhysRevResearch.7.013214},
  url = {https://link.aps.org/doi/10.1103/PhysRevResearch.7.013214}
}

@article{mataragkas25b,
  title = {{Transfer-matrix approach to the Blume-Capel model on the triangular lattice}},
  author = {Mataragkas, D and Vasilopoulos, A and Fytas, N G and Kim, D-H},
  journal = {Phys. Rev. Res.},
  volume = {7},
  issue = {3},
  pages = {033240},
  numpages = {10},
  year = {2025},
  month = {Sep},
  publisher = {American Physical Society},
  doi = {10.1103/jfl3-f4kd},
  url = {https://link.aps.org/doi/10.1103/jfl3-f4kd}
}

@article{fytas13b,
	author = {Fytas, N G and Theodorakis, P E},
	date = {2013/01/31},
	doi = {10.1140/epjb/e2012-30705-x},
	id = {Fytas2013},
	isbn = {1434-6036},
	journal = {The European Physical Journal B},
	number = {2},
	pages = {30},
	title = {{Universality aspects of the 2d random-bond Ising and 3d Blume-Capel models}},
	url = {https://doi.org/10.1140/epjb/e2012-30705-x},
	volume = {86},
	year = {2013}}

@article{hasenbusch10,
  title = {{Finite size scaling study of lattice models in the three-dimensional Ising universality class}},
  author = {Hasenbusch, M},
  journal = {Phys. Rev. B},
  volume = {82},
  issue = {17},
  pages = {174433},
  numpages = {13},
  year = {2010},
  month = {Nov},
  publisher = {American Physical Society},
  doi = {10.1103/PhysRevB.82.174433},
  url = {https://link.aps.org/doi/10.1103/PhysRevB.82.174433}
}

@article{moueddene24,
  title = {{Critical and tricritical behavior of the $d=3$ Blume-Capel model: Results from small-scale Monte Carlo simulations}},
  author = {Moueddene, L and Fytas, N G and Berche, B},
  journal = {Phys. Rev. E},
  volume = {110},
  issue = {6},
  pages = {064144},
  numpages = {12},
  year = {2024},
  month = {Dec},
  publisher = {American Physical Society},
  doi = {10.1103/PhysRevE.110.064144},
  url = {https://link.aps.org/doi/10.1103/PhysRevE.110.064144}
}

\end{document}